\documentclass[twocolumn,aps,floatfix,superscriptaddress,longbibliography]{revtex4-1}
\pdfoutput=1 
\usepackage{amsmath,amssymb,eucal,graphicx,float,epstopdf}

\usepackage[colorlinks=true, urlcolor=blue, anchorcolor=blue, citecolor=blue,filecolor=blue,linkcolor=blue,menucolor=blue]{hyperref}

\begin{document}
\title{Statistical properties of sites visited by independent random walks}
\author{E.~Ben-Naim}
\affiliation{Theoretical Division and Center for Nonlinear Studies, Los Alamos National Laboratory, Los Alamos, New Mexico 87545, USA}
\author{P.~L.~Krapivsky}
\affiliation{Department of Physics, Boston University, Boston, Massachusetts 02215, USA}
\affiliation{Santa Fe Institute, Santa Fe, New Mexico 87501, USA}

\begin{abstract}
The set of visited sites and the number of visited sites are two basic
properties of the random walk trajectory. We consider two independent
random walks on hyper-cubic lattices and study ordering probabilities
associated with these characteristics.  The first is the probability 
that during the time interval $(0,t)$, the number of sites visited by
a walker never exceeds that of another walker. The second is the
probability that the sites visited by a walker remain a subset of the
sites visited by another walker.  Using numerical simulations, we
investigate the leading asymptotic behaviors of the ordering
probabilities in spatial dimensions $d=1,2,3,4$.  We also study the
time evolution of the number of ties between the number of visited
sites.  We show analytically that the average number of ties increases
as $a_1\ln t$ with $a_1=0.970508$ in one dimension and as $(\ln t)^2$
in two dimensions.
\end{abstract}

\maketitle

\section{Introduction}

Random walk is an elementary random process which is ubiquitous in
several branches of mathematics, physics, chemistry, biology, finance,
etc. \cite{Chandrasekhar43,Feller,Berg93,Kampen,Money}. Questions
involving large deviations, persistence, and geometrical
characteristics of random walks continue to emerge
\cite{Chen10,Popov21}. Here, we investigate ordering probabilities
associated with the set of sites visited by independent random walks.

The maximum position attained by the walk is a basic characteristic of
the set of visited sites. The maxima of two one-dimensional random
walks remain ordered up to time $t$ with a probability that decays as
$t^{-1/4}$ \cite{BK-maxima, Julien-maxima}.  In general, it is
difficult to compute ``persistence'' exponents for non-Markovian
quantities such as the maximal position of a random walk
\cite{Bray13,Aurzada}. Nevertheless, the persistence exponent $1/4$
can be derived analytically \cite{BK-maxima, Julien-maxima}.  Further,
it can also be shown that the average number of lead changes $A(t)$
grows logarithmically with time \cite{BK-maxima-leads}
\begin{equation}
\label{ties-M}
A(t) \simeq \pi^{-1}\ln t
\end{equation}

The maximal position is (i) not uniquely defined in higher dimensions,
and (ii) does not \cite{Berkowitz,JML} necessarily increase by equal
amounts~\footnote{For instance, the maximal radial distance from the
  starting point is a piecewise constant function, but the lengths
  vary from jump to jump; the range of feasible jump lengths remains
  $\leq 1$ and increases with time.}. In this study, we focus on the
total number $\mathcal{N}(t)$ of distinct sites visited by a random
walk, {\it range} in short; in one dimension,
\hbox{$\mathcal{N}(t)=M(t)-m(t)+1$} where $M(t)$ is the maximum, and
$m(t)$ is the minimum. Unlike the maximum, the range is well-defined in
arbitrary dimension; moreover, it is a piecewise constant function of
time that increases by one.

We investigate the ``competition'' between the ranges of, as well as
the sets of sites visited by, two independent random walks.
Specifically, we consider two identical random walks with the same
starting position on hyper-cubic lattices $\mathbb{Z}^d$ in dimension
$d$. In each step, each random walk moves to one of its $2d$
neighboring sites, a site that is selected randomly and independently.
We study survival probabilities associated with the number of visited
sites and the set of visited sites in dimensions $d=1,2,3,4$. (We
expect that the asymptotic behavior for $d=4$ holds for all $d>4$.)
Our extensive numerical simulations reveal a diverse set of asymptotic
behaviors ranging from power laws and stretched exponentials to simple
exponentials. We also find asymptotic behaviors varying logarithmically with time.

\begin{figure}[t]
\begin{center}
\includegraphics[width=0.4\textwidth]{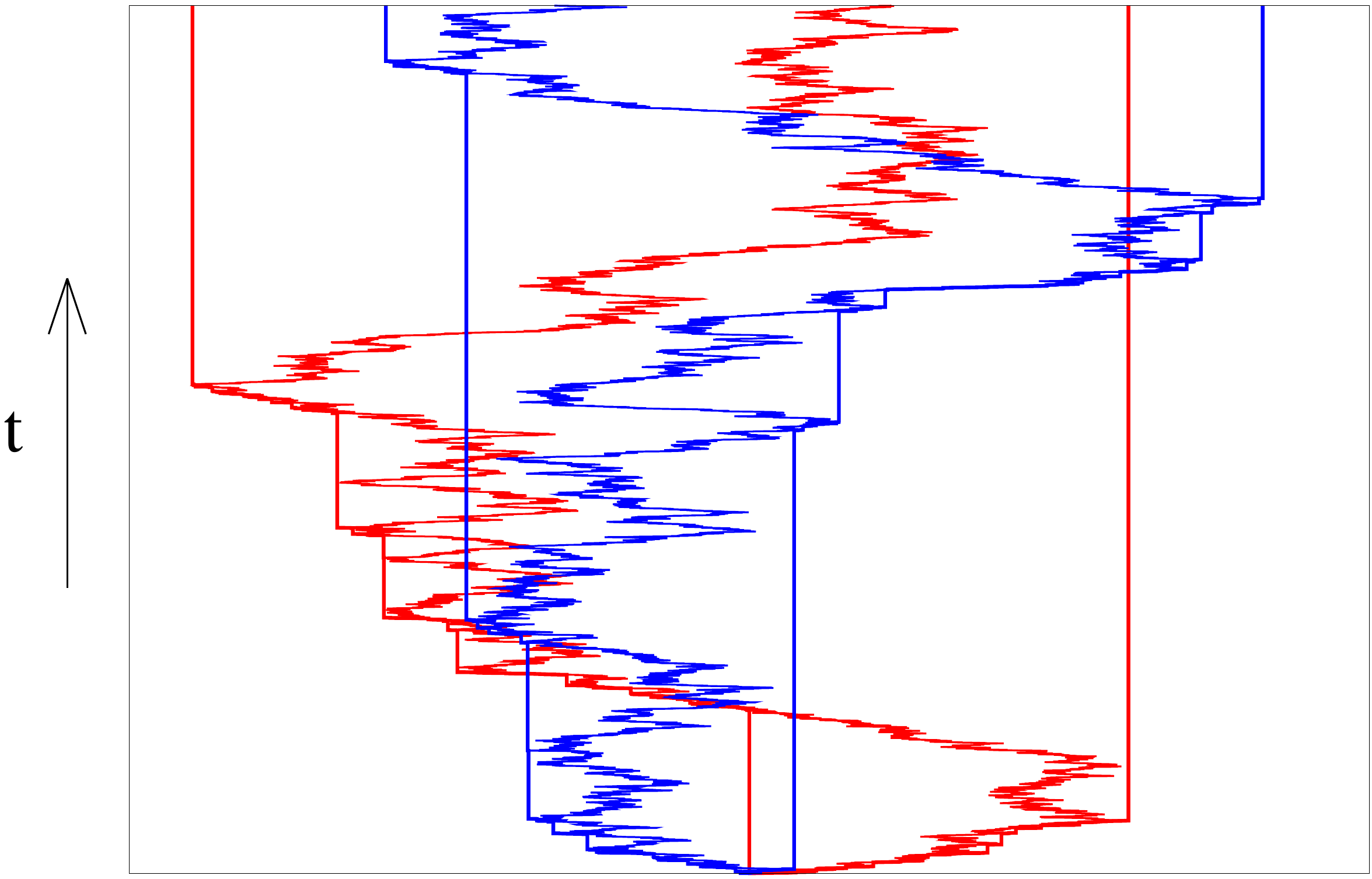}
\caption{Spacetime diagrams of two random walkers where the range of
  the random walker shown in red always exceeds the range of the
  random walker shown in blue.  The maximal and minimal positions of
  the two walkers are also indicated.}
\label{fig-range}
  \end{center}
\end{figure}

Let $\mathcal{N}_j(t)$ be the number of sites visited by the
$j^\text{th}$ walker: \hbox{$\mathcal{N}_j(t)=\mathcal{N}_j(t-1)+1$}
if at time $t$ the $j^\text{th}$ walker hops to a previously unvisited
site. Initially $\mathcal{N}_1(0)=\mathcal{N}_2(0)=1$. The ordering
probability associated with the ranges $\mathcal{N}_1$ and
$\mathcal{N}_2$ is [see Fig.~\ref{fig-range}]
\begin{equation}
\label{Pt:def}
P(t) = \text{Prob}[\mathcal{N}_1(\tau)\leq  \mathcal{N}_2(\tau)\,|\,0\leq \tau\leq t]
\end{equation}
In other words, $P(t)$ is the probability that a random walker never
visits more sites than another independent random walker up time $t$.
The random quantities $\mathcal{N}_1$ and $\mathcal{N}_2$ are
independent and non-Markovian, and this feature makes determination of
the ordering probability $P(t)$ challenging \cite{Bray13,Aurzada}.

\begin{figure}[t]
\begin{center}
\includegraphics[width=0.4\textwidth]{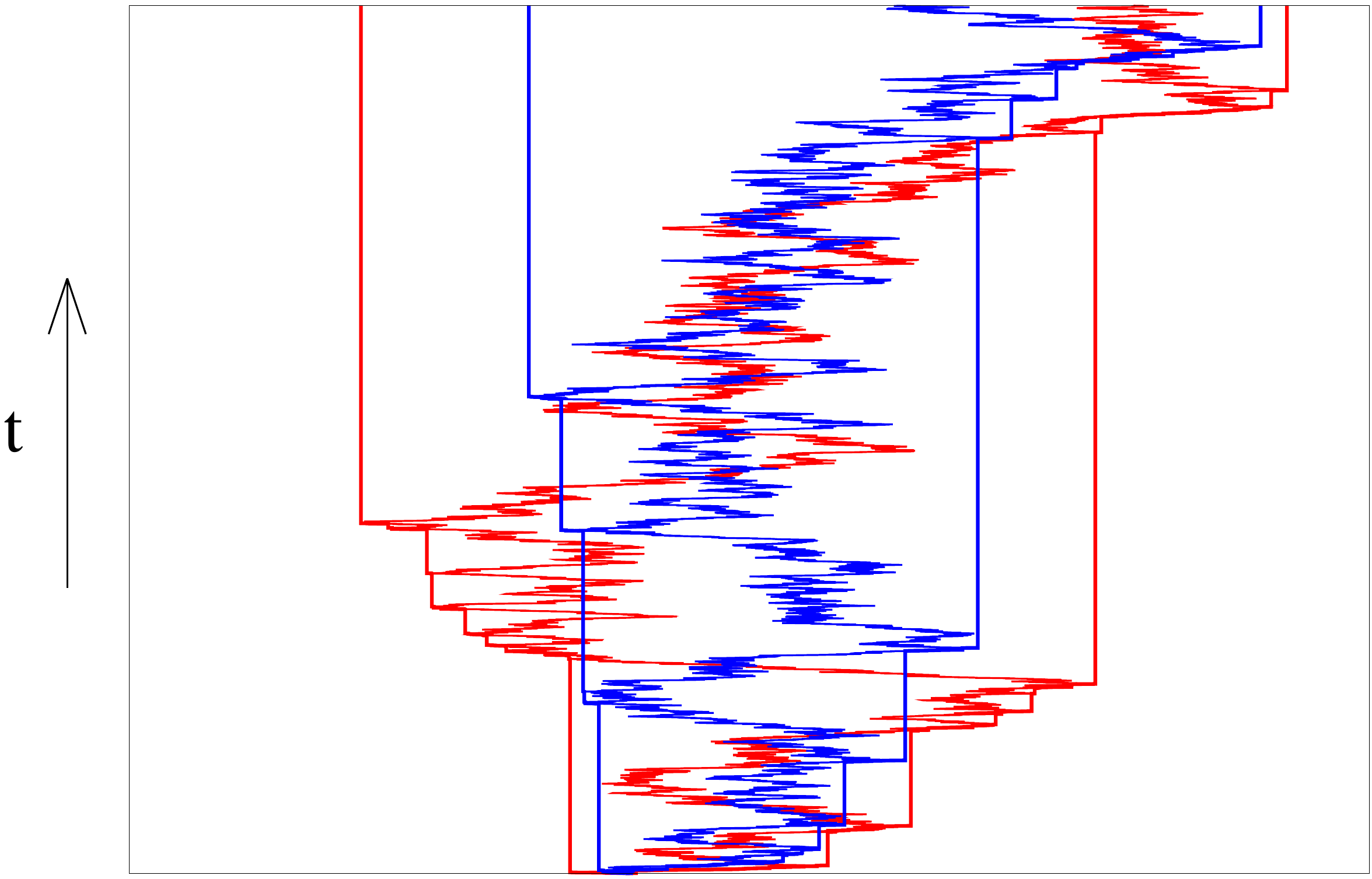}
\caption{Spacetime diagrams of two random walkers where each site
  visited by the random walker shown in blue has been previously
  visited by the random walker shown in red.  The maximal and minimal
  positions of both walkers are also displayed.}
\label{fig-shadow}
  \end{center}
\end{figure}

One can also compare the sets of sites visited by the two walkers,
denoted by $S_1(t)$ and $S_2(t)$.  For the sets of visited sites, the
natural ordering is inclusion~\footnote{In contrast to the linear
  order between, e.g., integers (any two integers are comparable),
  inclusion is a partial order: Two sets $U$ and $V$ can be either
  comparable ($U\subset V$, or $U=V$, or $V\subset U$), or $U$ and $V$
  can be incomparable.}.  The ordering probability associated with the
sets $S_1(t)$ and $S_2(t)$ is
\begin{equation}
\label{Ot:def}
Q(t) = \text{Prob}[S_1(\tau)\subseteq S_2(\tau)\,|\,0\leq \tau\leq t]
\end{equation}
Hence, $Q(t)$ is the probability that a walker never visits a site
that has not been previously visited by another independent walker up to time
$t$.  One can visualize this condition as a ``matryoshka'' arrangement
with the set $S_1$ always remaining a subset of $S_2$ throughout the
time interval $(0,t)$, see Fig.~\ref{fig-shadow}.

Since $\mathcal{N}_j = |S_j|$, where $|S|$ denotes the number of
elements in set $S$, the probability $Q(t)$ is bounded from above by
$P(t)$:
\begin{equation}
\label{bound}
Q(t) \leq P(t) 
\end{equation}
for all $t\geq 0$. Our simulations show that the ordering
probabilities $P(t)$ and $Q(t)$ decay algebraically in one dimension
\begin{align}
\label{Probs:1d}
P(t)\sim t^{-\beta}, \qquad  Q(t) \sim t^{-\gamma}
\end{align}
with $\beta=0.667\pm 0.002$ and $\gamma=1.45\pm 0.03$. The algebraic
decays \eqref{Probs:1d} are consistent with the asymptotic behavior of
the ordering probability associated with maxima
\cite{BK-maxima,Julien-maxima}.

We also study the number of distinct ties, that is, instances when
$\mathcal{N}_1$ equals $\mathcal{N}_2$ and vice versa. Our theoretical
results suggest that the average number of ties during the time
interval $(0,t)$ grows as
\begin{equation}
\label{ties-d}
A(t) \simeq
\begin{cases}
a_1\, \ln t                             & d=1\\
a_2\, (\ln t)^2                       & d=2\\
a_3\, t^{1/2}(\ln t)^{-1/2}      & d= 3\\
a_d\,  t^{1/2}                        & d\geq 4
\end{cases}
\end{equation}

In Sec.~\ref{sec:single}, we recall a few basic results about
statistics of the range of a random walk.  In Sec.~\ref{sec:order} we
present the asymptotic behaviors of the ordering probabilities $P(t)$
and $Q(t)$ suggested by numerical simulations, and we also provide
heuristic arguments supporting some of these behaviors.  In
Sec.~\ref{sec:ties}, we study the average number of ties and obtain
the growth laws \eqref{ties-d} theoretically. Generalizations to
multiple independent random walks are outlined in
Sec.~\ref{sec:many}. We conclude with a discussion
(Sec.~\ref{sec:disc}).

\section{Range of a random walk}
\label{sec:single}

The number of distinct sites visited by a random walk, namely the
range, has been the subject of considerable research
\cite{Polya21,Polya38,Dvoretzky,Weiss65,Weiss}. Statistical properties
of the range are well understood in one dimension
\cite{Daniels,Kuhn,Feller51,Rubin}, but remain incomplete in higher
dimensions
\cite{Jain68,Jain71,Jain72,Jain74,Weiss-Rubin,LeGall86,LeGall91,Hamana97,Hughes,Asselah18,Asselah19,Sebek}.

We now summarize key statistical properties of the range, which we
later use to analyze the growth laws \eqref{ties-d}. These results
apply to a symmetric nearest-neighbor random walk on the hyper-cubic lattice
$\mathbb{Z}^d$ in dimension $d$. The overall hopping rate is set to
unity so that the variance in the displacement ${\bf r}$ equals time,
$\langle {\bf r}^2\rangle -\langle {\bf r}\rangle^2=t$. The leading
asymptotic behaviors of the average range $N(t)=\langle
\mathcal{N}(t)\rangle$ are
\begin{equation}
\label{Nt-d}
N(t)\simeq
\begin{cases}
\sqrt{\frac{8t}{\pi}}      & d=1\\
\frac{\pi t}{\ln t}          & d=2\\
t/W_d                        & d>2
\end{cases}
\end{equation}
where $W_d$ are the so-called Watson integrals
\cite{Watson,Glasser77,Guttmann10,Zucker11}. For hyper-cubic
lattices 
\begin{equation}
\label{Watson}
W_d=\int_0^{2\pi}\ldots\int_0^{2\pi} \left[1-\frac{1}{d}\sum_{i=1}^d \cos q_i\right]^{-1}\prod_{i=1}^d \frac{dq_i}{2\pi}
\end{equation} 
when $d\geq 3$. For the cubic lattice, the Watson integral can be 
expressed \cite{Glasser77} via the gamma function,
\begin{eqnarray}
\label{W3}
W_3 = \frac{\sqrt{6}}{32\,\pi^3}\, 
\Gamma\left(\frac{1}{24}\right)\, \Gamma\left(\frac{5}{24}\right)\, 
\Gamma\left(\frac{7}{24}\right)\, \Gamma\left(\frac{11}{24}\right)
\end{eqnarray} 

The asymptotic behavior of the variance, \hbox{$V=\langle
\mathcal{N}^2\rangle-\langle \mathcal{N}\rangle^2$}, is also known
\begin{equation}
\label{Vt-d}
V(t) \simeq
\begin{cases}
(4\ln 2-8/\pi)\,t                 & d=1\\
V_2\, t^2/(\ln t)^4                & d=2\\
V_3\, t \ln t                     & d=3\\
V_d\, t                           & d\geq 4
\end{cases}
\end{equation}
The amplitudes for square and cubic lattices are
\begin{equation*}
\begin{split}
V_2 & = \pi^2-\frac{\pi^4}{6}-2\pi^2\int_0^1dx\,\frac{\ln x}{1-x+x^2}=16.768\,193\ldots \\
V_3 & = 4 \pi^{-2}(1-1/W_3)^4 = 0.005\,450\,284\ldots
\end{split}
\end{equation*}
where $W_3$ is given by \eqref{W3}, see
\cite{Jain71,Jain72,Jain74,Hughes} for derivations of the amplitudes
$V_2$ and $V_3$. However, no compact formulas are available for the
amplitude $V_d$ when $d\geq 4$.

Equations \eqref{Nt-d} and \eqref{Vt-d} imply that the random quantity
$\mathcal{N}(t)$ is non-self-averaging in one dimension and
self-averaging when $d\geq 2$. Further, $\mathcal{N}(t)$ is weakly
self-averaging in two dimensions since the ratio $\sqrt{V}/N$ vanishes
very slowly as $(\ln t)^{-1}$.

The random variable $\mathcal{N}$ is fully characterized by the
distribution $P_n(t)=\text{Prob}[\mathcal{N}(t)=n]$.  In one
dimension, the range distribution converges to
\cite{Daniels,Kuhn,Feller51,Rubin}
\begin{equation}
\label{Pnt-1}
P_n(t)\simeq \frac{8}{\sqrt{2\pi t}}\sum_{j\geq 1}(-1)^{j-1} j^2\,\exp\!\left[-\tfrac{j^2 n^2}{2t}\right]
\end{equation}
For a random walk on a square lattice, the range distribution is
non-Gaussian \cite{LeGall86}, and a closed explicit expression for the
asymptotic range distribution remains elusive.  When $d\geq 3$, the
range distribution is asymptotically Gaussian
\cite{Jain71,Jain72,Jain74}
\begin{equation}
\label{Pnt-d}
P_n(t)\simeq \frac{1}{\sqrt{2\pi V(t)}}\,\exp\!\left\{-\frac{[n-N(t)]^2}{2V(t)}\right\}
\end{equation}
with $N(t)$ and $V(t)$ given by \eqref{Nt-d} and \eqref{Vt-d}. 

\begin{figure}
\centering
\includegraphics[width=0.4\textwidth]{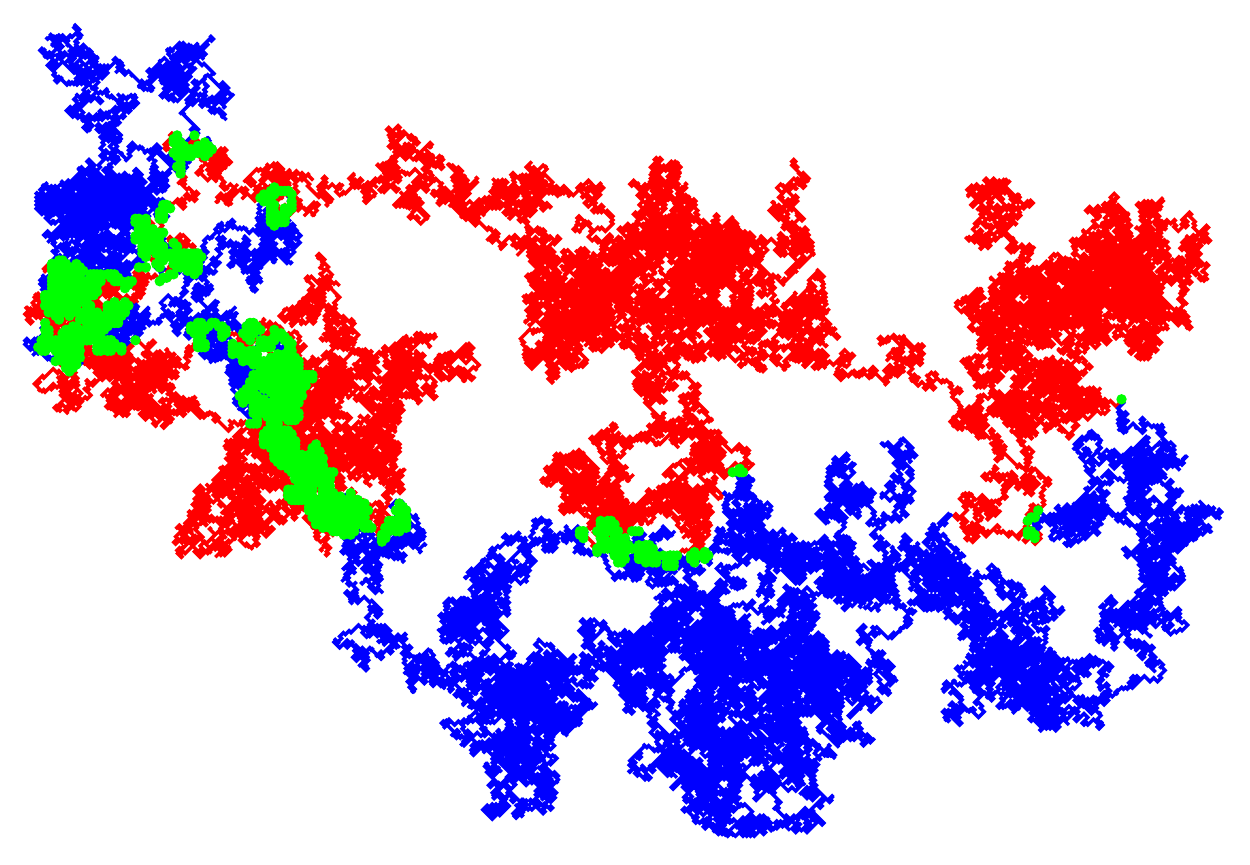}
\caption{An illustration of sites visited by two independent random
  walkers on the square lattice after $10^5$ steps.  Sites visited
  only by the first walker are shown in red, sites visited only by the
  second walker are shown in blue, and sites visited by both walkers
  are shown in green.}
\label{Fig:2d-common}
\end{figure}

The number of common sites $\mathcal{C}=|S_1 \cap S_2|$ quantifies the
overlap between the sites visited by two independent random walkers.
The average number of common sites, $C(t)=\langle
\mathcal{C}(t)\rangle$, grows according to
\begin{equation}
\label{Ct-d}
C(t) \sim
\begin{cases}
t^{1/2}              & d = 1\\
t/(\ln t)^2          & d = 2\\
t^{1/2}              & d = 3\\
\ln t                  & d = 4\\
1                      & d \geq 5
\end{cases}
\end{equation}
See \cite{Tamm12,KMS} for the derivation of \eqref{Ct-d} and
generalization to common sites visited by $m$ walkers with arbitrary
$m$. Figure \ref{Fig:2d-common} illustrates the number of common sites
visited by two walkers on the square lattice.

\section{Ordering Probabilities}
\label{sec:order}

Here, we analyze the evolution of the ordering probabilities $P(t)$
and $Q(t)$ using numerical simulations.  We implement the random walk
process in the standard way:
\begin{enumerate}
\item Initially, the random walk is at the origin. 
\item At each time step, the walker hops to one of its $2d$
  neighboring sites, a site that is chosen at random. Therefore, 
  throughout the evolution, the average displacement remains equal to
  zero. 
\item Time is augmented by one after each step.
\end{enumerate}
With this implementation, the variance of the displacement ${\bf
  r}(t)$ equals time, $\langle {\bf r}^2(t)\rangle=t$.

In one dimension, we keep track of three quantities: the current
position of the walk, the leftmost position $m(t)$ and the rightmost
position $M(t)$; the total number of visited sites is given by ${\cal
  N}(t)=M(t)-m(t)+1$. Hence, the required computer memory is minimal.
In higher dimensions, it is necessary to maintain a physical lattice
to indicate which sites were visited by the walker and which remain
unvisited. The simulations can be still performed efficiently by
keeping track of all sites visited by the walk and resetting {\em
  only} the visited sites on the indicator lattice at the end of each
run.  This approach is especially well suited for measuring 
survival probabilities.

\begin{figure}[t]
\begin{center}
\includegraphics[width=0.425\textwidth]{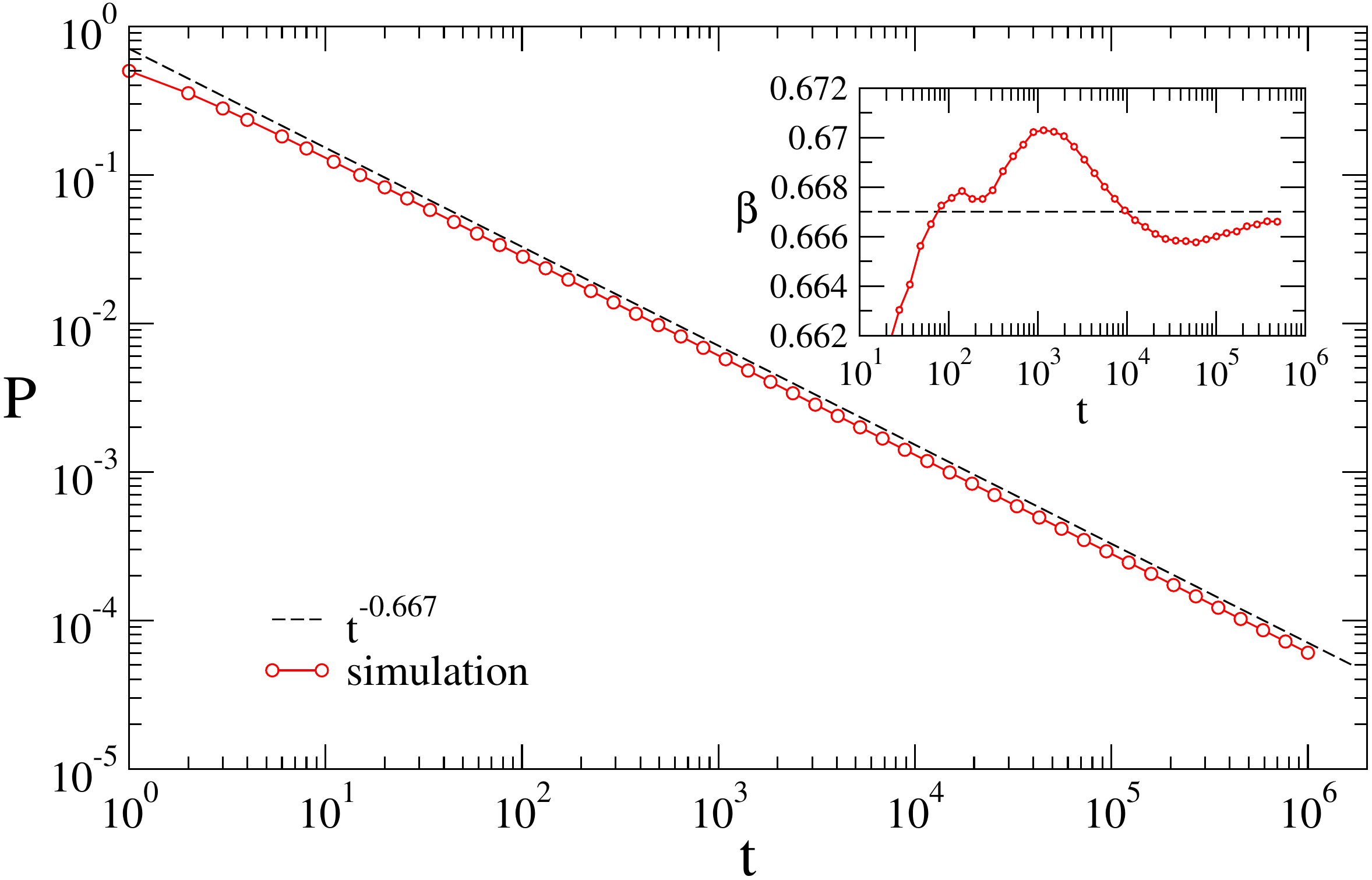}
\caption{The ordering probability $P(t)$ versus time $t$ in one
  dimension. An average over $2^{38}$ independent runs has been
  performed. Also shown for reference is a line with slope
  $0.667$. The inset shows the local slope $\beta\equiv -d \ln P/d \ln
  t$.}
\label{fig-1d-P}
\end{center}
\end{figure}

\begin{figure}[t]
\begin{center}
\includegraphics[width=0.425\textwidth]{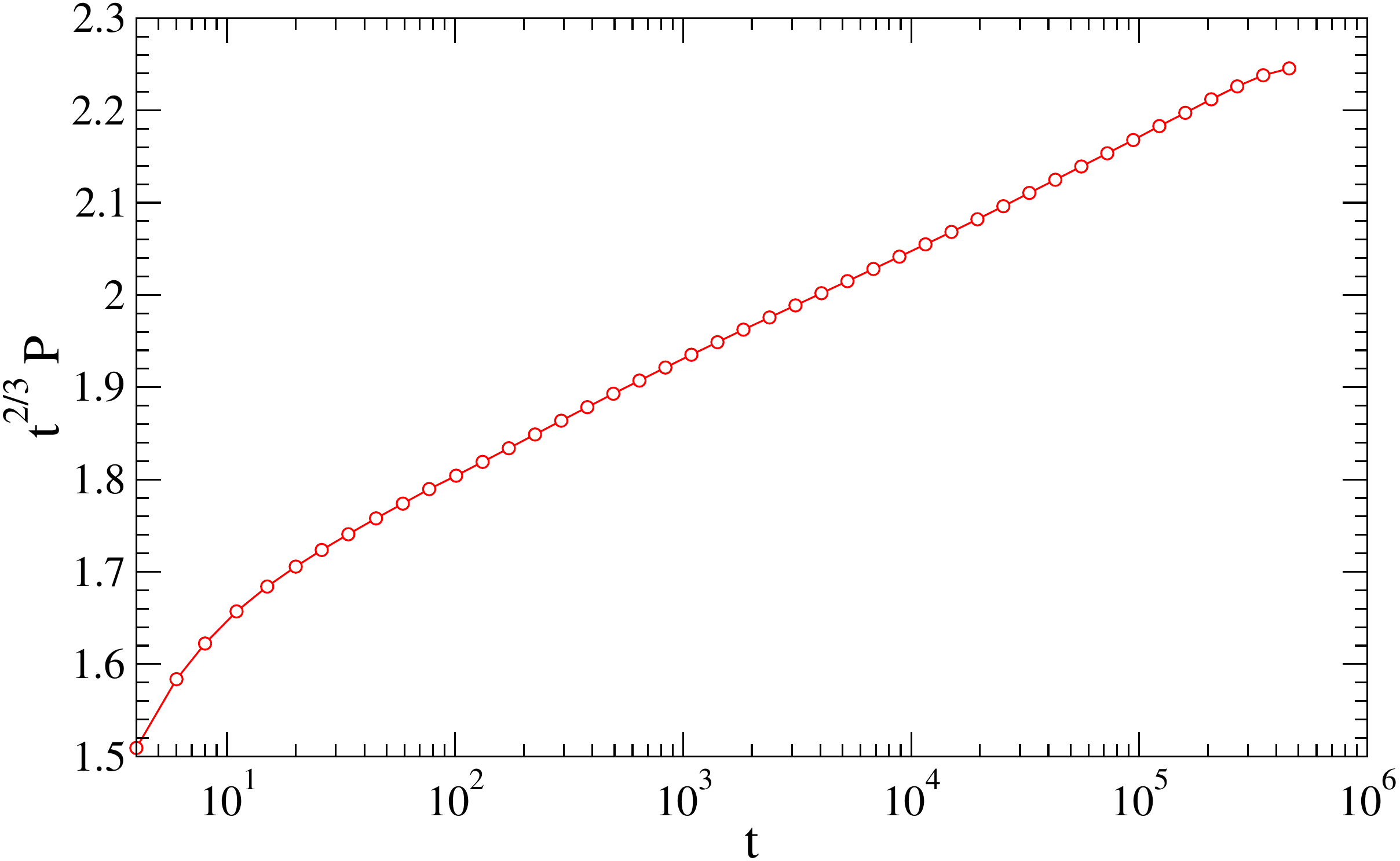}
\caption{The quantity $t^{2/3}P(t)$ versus time $t$ on a square
  lattice. Simulation results represent an average of $2^{34}$
  independent realizations.}
\label{fig-2d-P}
\end{center}
\end{figure}

\begin{figure}[t]
\begin{center}
\includegraphics[width=0.425\textwidth]{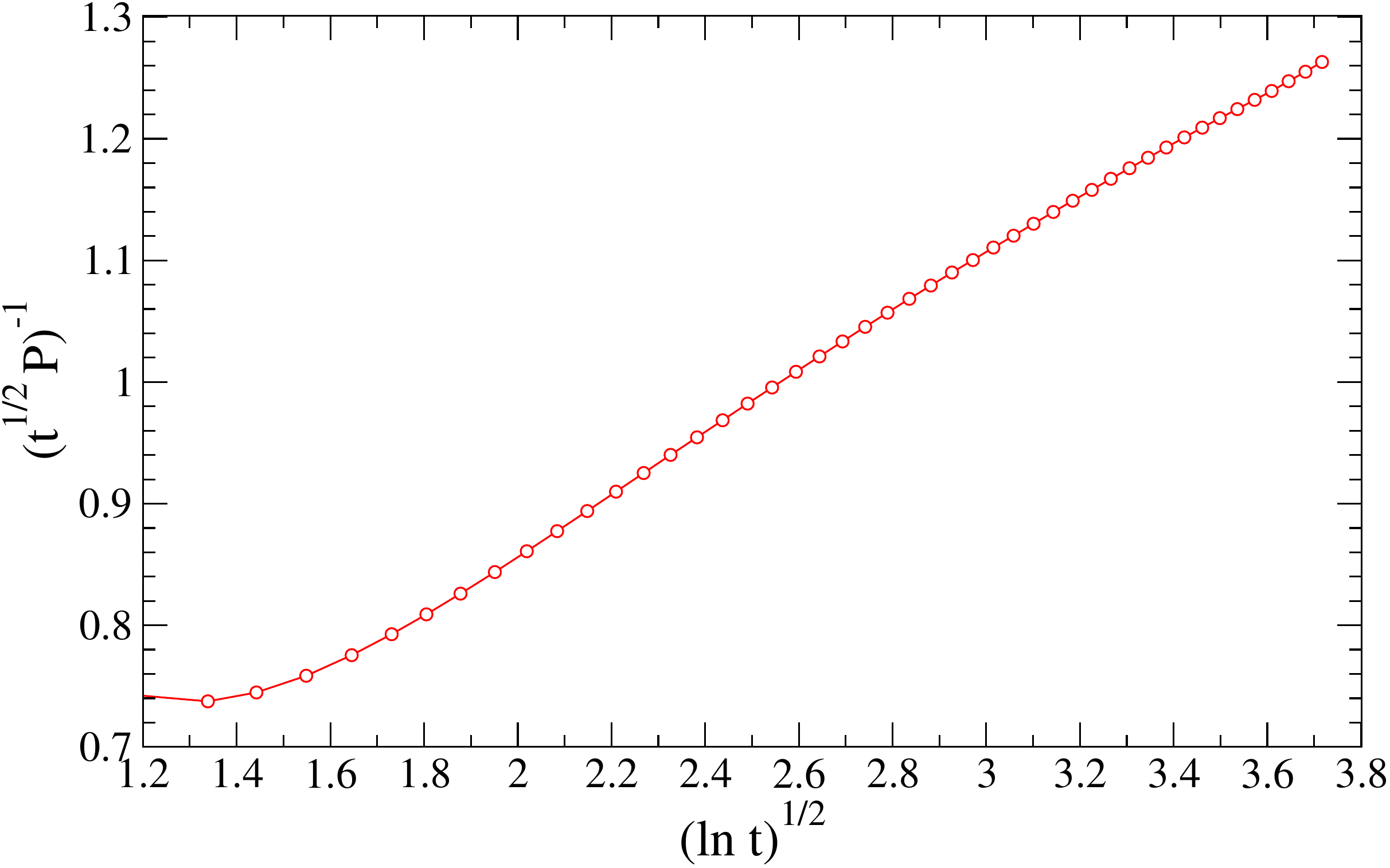}
\caption{The quantity $[t^{1/2}P(t)]^{-1}$ versus $\sqrt{\ln t}$ on a
  cubic lattice. Simulation results represent an average of
  $2^{30}$ runs.}
\label{fig-3d-P}
\end{center}
\end{figure}

The Monte Carlo simulation results suggest the following asymptotic
behaviors for the ordering probability $P(t)$ associated with the
range \eqref{Pt:def} [see Figs.~\ref{fig-1d-P}--\ref{fig-3d-P}]
\begin{equation}
\label{Pt-d-precise}
P(t) \sim
\begin{cases}
t^{-\beta}                            & d=1\\
t^{-\beta}\, \ln t                   & d=2\\
t^{-1/2}(\ln t)^{-1/2}            & d=3\\
t^{-1/2}                              & d= 4
\end{cases}
\end{equation}
Simulations suggest a simple rational value $\beta=2/3$
for the persistence exponent; specifically, we measure
\begin{equation}
\label{alpha}
\beta = 0.667\pm 0.002
\end{equation}
in one dimension (see Fig.~\ref{fig-1d-P}).  In two dimensions, the
effective exponent is only slightly smaller than $2/3$, and moreover,
the quantity $-d \ln P/d \ln t$ increases slowly with time. These
observations indicate a possible logarithmic correction, and indeed,
the simulations support the decay $P\sim t^{-2/3} \ln t$, see
Fig.~\ref{fig-2d-P}. A simple $t^{-1/2}$ decay emerges in four
dimensions, and we expect this behavior extends to $d> 4$.  In three
dimensions, the effective exponent is slightly larger than
$1/2$. Moreover, the quantity $-d \ln P/d \ln t$ decreases slowly with
time, again indicating a logarithmic correction.  The numerical
results support the decay law $P\sim t^{-1/2}(\ln t)^{-1/2}$, see
Fig.~\ref{fig-3d-P}.

When $d\geq 4$, the ranges $\mathcal{N}_1(t)$ and $\mathcal{N}_2(t)$
perform independent directed random walks, so $P\sim t^{-1/2}$. The 
logarithmic correction to the $t^{-1/2}$ asymptotic in three
dimensions is due to the temporal behavior of the variance, see
Eq.~\eqref{Vt-d}. Comparing Eq.~\eqref{Pt-d-precise} with the variance
in the number of visited sites, Eq.~\eqref{Vt-d}, we observe that
$P\propto V^{-1/2}$ when $d\geq 3$. The first-passage probability for
a broad class of one-dimensional Markovian random variables decays as
$(\text{variance})^{-1/2}$, see \cite{Novikov83,Novikov87,Denisov}. Therefore, the first-passage
probability for the random variable $\mathcal{N}_1-\mathcal{N}_2$,
namely $P(t)$, is expected to have this property when $d\geq 3$ as
$\mathcal{N}_1$ and $\mathcal{N}_2$ become uncorrelated and Markovian
in the asymptotic limit.

In one dimension, the algebraic decay of $P(t)$ is consistent with the
behavior found for the maxima \cite{BK-maxima}.  It would be
interesting to find a heuristic explanation for the logarithmic
enhancement of the ordering probability $P(t)$ in two dimensions. We
note that logarithmic terms arise in Eqs.~\eqref{Nt-d}, \eqref{Vt-d},
and also characterize the support of the two-dimensional random walk,
see \cite{Dvoretzky50,Taylor60a,Taylor60b,Hilhorst91,Hilhorst93,Hilhorst96,Hilhorst97,Peres01,Peres06,BM:book}.

The ordering probability associated with the set of visited sites
$Q(t)$ is bounded from above by the quantity $P(t)$, as stated in
Eq.~\eqref{bound}. Indeed, the condition in \eqref{Ot:def} is
significantly more stringent than the condition in \eqref{Pt:def} . In
one dimension
\begin{align}
\label{gamma}
Q(t) \sim t^{-\gamma}\,, \quad \gamma = 1.45 \pm 0.03
\end{align}
Both $P(t)$ and $Q(t)$ decay algebraically in one dimension. The
inequality $\gamma>\beta$ follows from Eq.~\eqref{bound}, yet a more
stringent relation, $\gamma> 1> \beta$, holds. These two inequalities
imply that average first-passage time associated with the survival
probability $P(t)$ is infinite, while that associated with the
quantity $Q(t)$ is finite.

\begin{figure}[t]
\begin{center}
\includegraphics[width=0.425\textwidth]{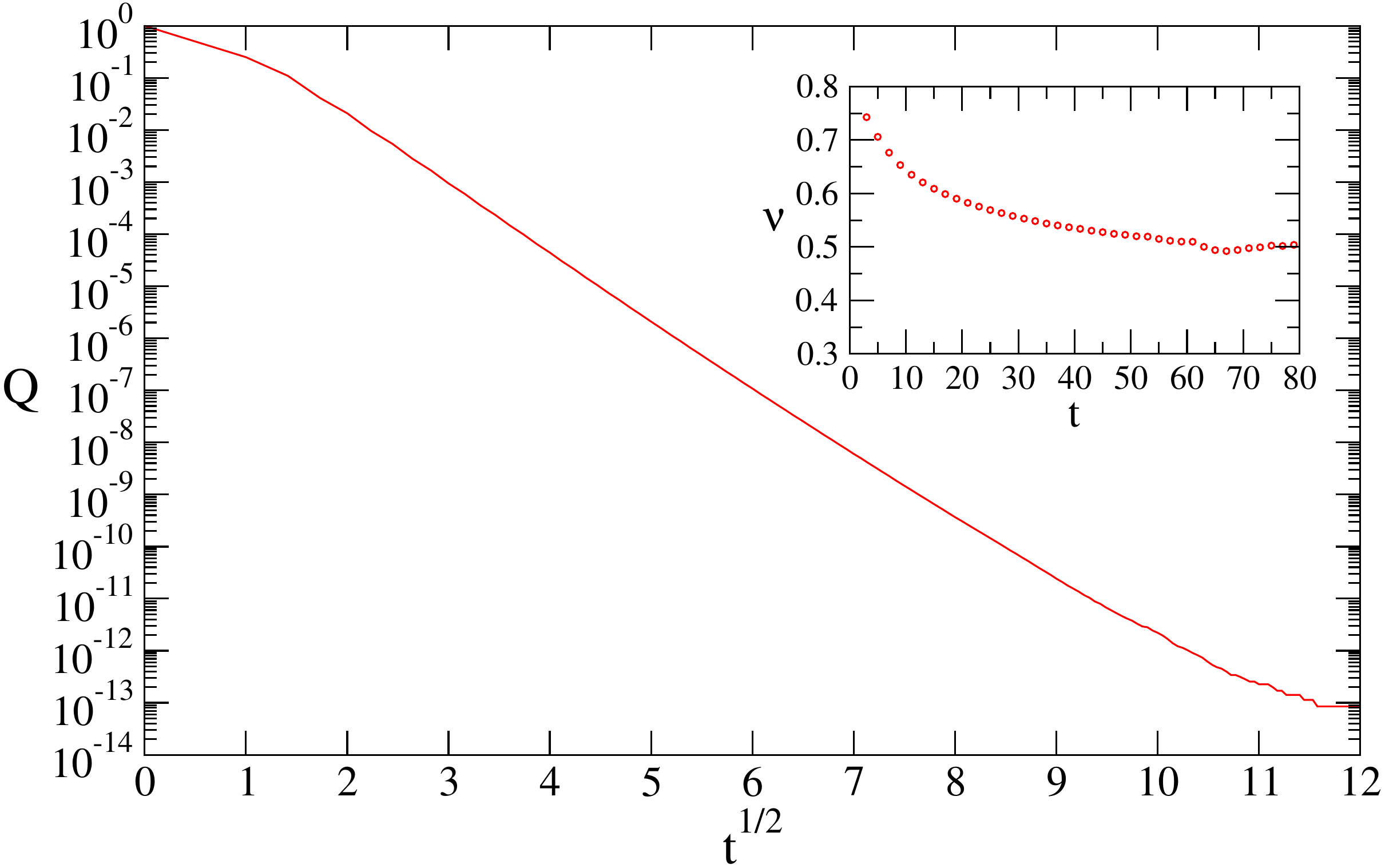}
\caption{A semi-log plot of $Q(t)$ versus $t^{1/2}$ in two
  dimensions. Simulation results represent an average of $2^{45}$
  independent runs. The inset displays the quantity $\nu \equiv d \ln
  \ln [1/Q]/ d \ln t$ versus time $t$.  }
\label{fig-2d-O}
\end{center}
\end{figure}

\begin{figure}[t]
\begin{center}
\includegraphics[width=0.425\textwidth]{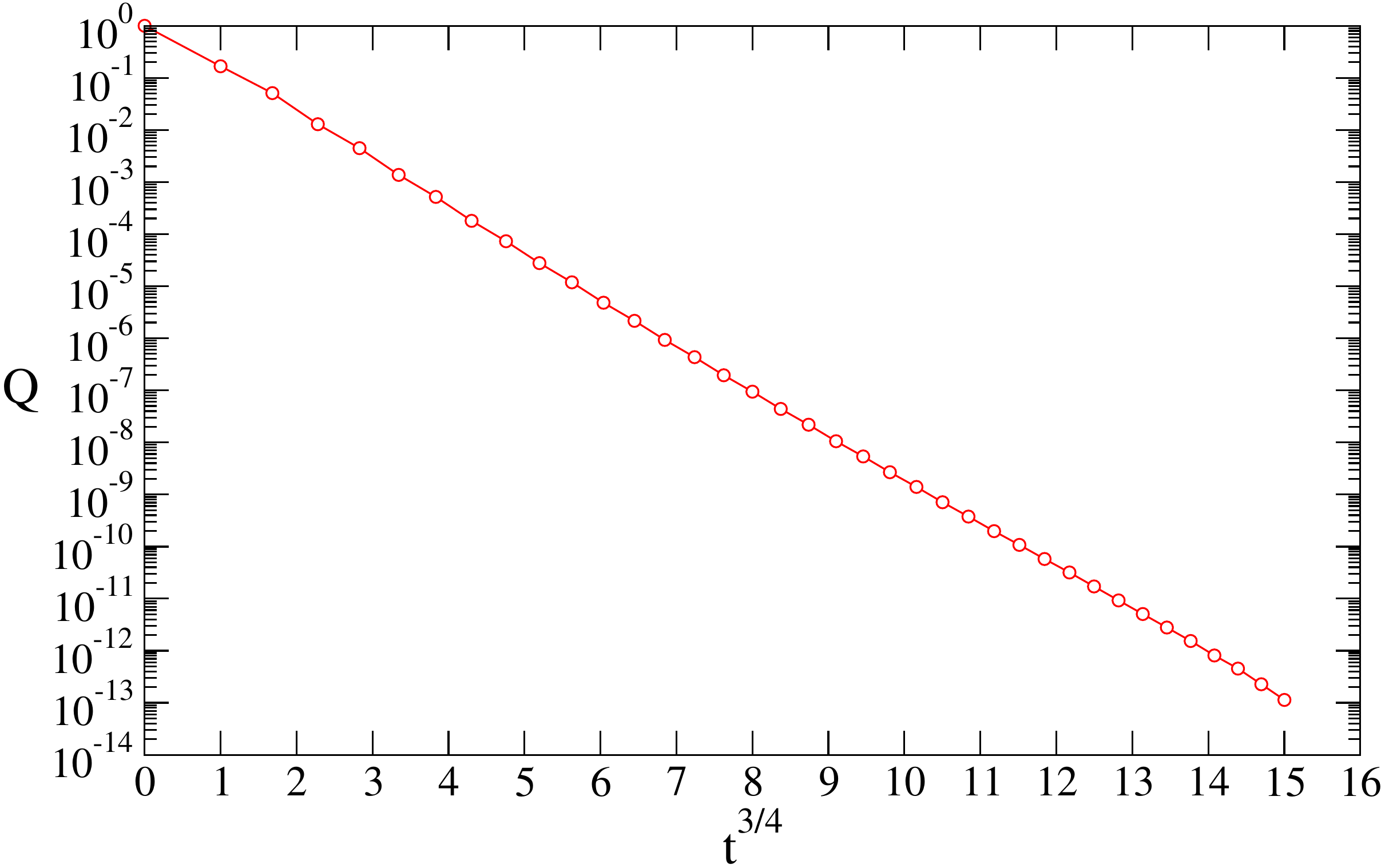}
\caption{A semi-log plot of $Q(t)$ versus $t^{3/4}$ in three
  dimensions. Simulation results represent an average of $2^{46}$
  independent runs.}
\label{fig-3d-O}
\end{center}
\end{figure}

\begin{figure}[t]
\begin{center}
\includegraphics[width=0.425\textwidth]{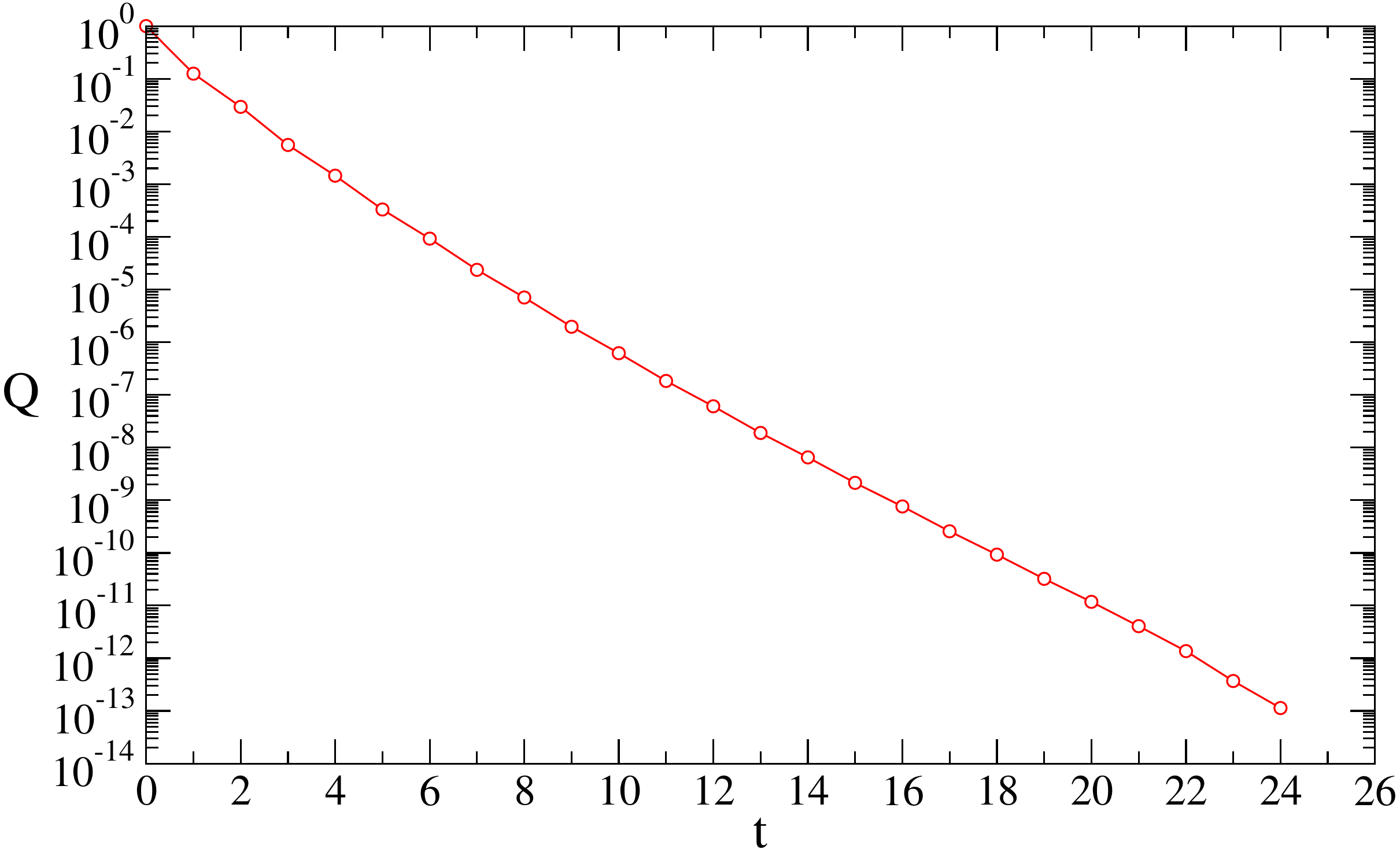}
\caption{A semi-log plot of $Q(t)$ versus $t$ in four
  dimensions. An average over $2^{45}$
  independent runs has been performed.}
\label{fig-4d-O}
\end{center}
\end{figure}

When $d\geq 2$, the ordering probability $Q(t)$ decays faster than any
power law. Numerical simulations
(Figs.~\ref{fig-2d-O}--\ref{fig-4d-O}) support stretched exponential
behaviors:
 \begin{equation}
\label{Ot-d-sim}
\ln[1/Q(t)] \sim
\begin{cases}
t^{1/2}         & d=2\\
t^{3/4}         & d= 3\\
t                  & d \geq  4
\end{cases}
\end{equation}
The temporal range probed by the simulations is much larger in
dimension $d=2$ compared with that in dimensions $d=3$ and $d=4$. In
$d=2$, the simulation results provide evidence in support of the
stretched exponential decay in \eqref{Ot-d-sim} as the quantity $\nu
\equiv d \ln \ln [1/Q(t)] /d\ln t$ saturates at the value $\nu=0.50\pm
0.01$; see inset to Fig.~\ref{fig-2d-O}, the region $t>80$ was excluded because of poor statistics.  
The sharp decay of the quantity $Q(t)$ in $d=3$ and $d=4$ dimensions makes it difficult to assess the
asymptotic behavior and the decays stated in \eqref{Ot-d-sim}
represent our best estimates based on extensive Monte Carlo
simulations. For instance, in $d=3$, the asymptotic behavior quoted in
\eqref{Ot-d-sim} is only slightly better aligned with the simulation
results than the decay $\ln [1/Q(t)]\sim -t/\ln t$.

Establishing the asymptotic behaviors stated in \eqref{Ot-d-sim}
theoretically is a formidable challenge.  In $d\geq 3$, a random walk
hops to an unvisited site with a non-vanishing probability in the
long-time limit, suggesting $Q(t)$ decays exponentially~\footnote{In
  three and higher dimensions, the random walker hops to an unvisited
  site with probability approaching $1/W_d$ as $t\to\infty$. This
  explains the asymptotic growth \eqref{Nt-d} of the average range,
  $N\simeq t/W_d$.}. However, we find purely exponential decay only
when $d\geq 4$.

To appreciate the  plausibility of the exponential decay, consider two random walks with identical trajectories. In this
scenario, the sets of sites visited by the two walkers are identical,
$S_1\equiv S_2$. Such a time evolution is realized with
probability $(2d)^{-t}$. This argument provides the lower bound $Q(t)\geq (2d)^{-t}$
and the upper bound $b_d\leq \ln (2d)$.  In $d=4$, the numerical
simulations give $b_4=1.05\pm 0.05$, while the upper bound is $\ln
8=2.079$.  We stress that the bound $b_d\leq \ln (2d)$ applies to a
discrete-time random walk.

\section{The number of ties}
\label{sec:ties}

We now study ties between the ranges $\mathcal{N}_1=|S_1|$ and $\mathcal{N}_2=|S_2|$, 
i.e., instances when the number of sites visited by the two random
walkers become equal.  The random
quantity $\mathcal{N}_1-\mathcal{N}_2$ is piecewise constant, and
changes by unit increments or decrements.  Specifically, we are
interested in {\em distinct} ties that occur when
$\mathcal{N}_1-\mathcal{N}_2$ resets to zero.  Let $T(t)$ be the
number of distinct ties during the time interval $(0,t)$. The initial
condition is $T(0)=1$.  We define $\Phi_n(t)=\text{Prob}[T(t)=n]$ to be
the probability the number of distinct ties at time $t$ equals
$n$. Our focus is the average number of ties,
\begin{equation}
\label{At:def}
A(t) = \sum_{n\geq 1}n\Phi_n(t)
\end{equation}

To obtain the asymptotic behavior of the average number of ties, we
use the general formula \cite{BK-maxima-leads}
\begin{equation}
\label{nu-PQ}
\frac{dA}{dt}= 2\sum_{n\geq 2} P_n \,\frac{d\mathbb{P}_n}{dt}
\end{equation}
This equation relates the growth of the average number of ties to the
range distribution $P_n$ and its corresponding cumulative distribution
\begin{equation}
\label{Q:def}
\mathbb{P}_n = \sum_{k\geq n} P_k
\end{equation}
To derive \eqref{nu-PQ} we note that $\mathcal{N}_1$ and
$\mathcal{N}_2$ are independent variables.  The number of ties can
increase only when: (i) these quantities differ by one, say
$\mathcal{N}_1=n-1$ and $\mathcal{N}_2=n$, and (ii) the smaller
quantity $\mathcal{N}_1$ increases, $n-1\to n$.  The factor $2$ in
\eqref{nu-PQ} accounts for the fact that either random walk may be in
the lead.  The rate by which the trailing walker makes the jump
$n-1\to n$, denoted by $W_{n-1,n}$, is the gain term in $\frac{d
  P_n}{dt}$, viz.
\begin{equation}
\label{Pnn}
\frac{d P_n}{dt} = W_{n-1,n} - W_{n,n+1}
\end{equation}
Similarly, we have 
\begin{equation}
\label{Pn+}
\begin{split}
\frac{d P_{n+1}}{dt} & = W_{n,n+1} - W_{n+1,n+2},\\
\frac{d P_{n+2}}{dt} & = W_{n+1,n+2} - W_{n+2,n+3},\\
\frac{d P_{n+3}}{dt} & = W_{n+2,n+3} - W_{n+3,n+4}, 
\end{split}
\end{equation}
etc. By summing \eqref{Pnn} and all the successive equations
\eqref{Pn+} we obtain
\begin{equation}
\frac{d\mathbb{P}_n}{dt} = \sum_{k\geq n} \frac{d P_k}{dt} = W_{n-1,n}
\end{equation}
thereby leading to \eqref{nu-PQ}. The rate equation \eqref{nu-PQ},
which utilizes continuous time, is suitable for describing the
long-time asymptotic behavior.

\begin{figure}[t]
\begin{center}
\includegraphics[width=0.425\textwidth]{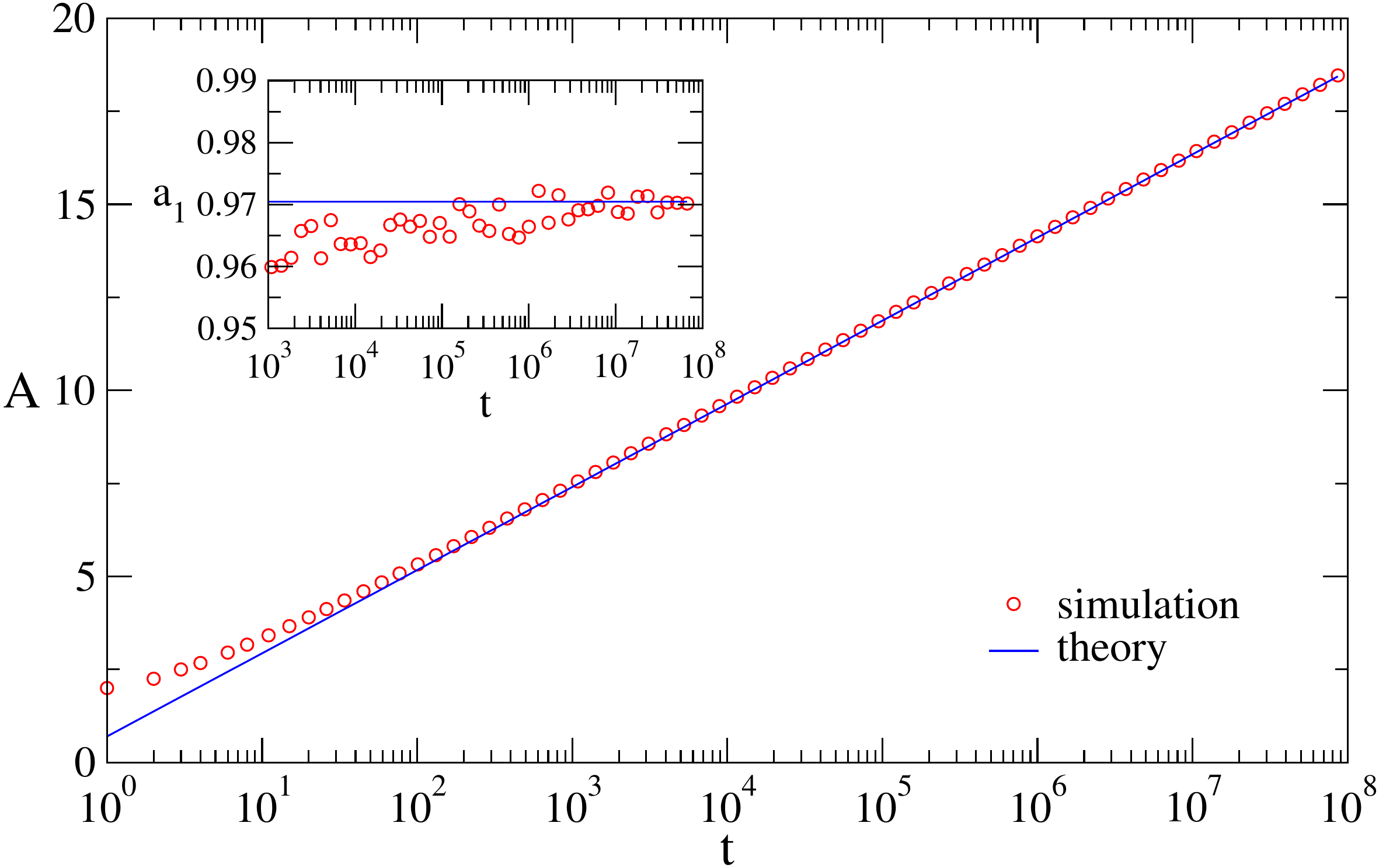}
\caption{The average number of ties $A$ versus time $t$ in one
  dimension. We compare simulation results with theory. Fitting to the
  form $A\simeq a_1\ln t$ in the range $10^6<t<10^8$ yields the
  estimate $a_1=0.970\pm 0.001$.  The inset compares simulation
  results for the quantity $a_1(t) \equiv dA/d\ln t$ with theoretical
  prediction, Eq.~\eqref{a1-simple}. Simulation results represent an
  average of $2^{20}$ independent runs.}
\label{fig-1d-nu}
  \end{center}
\end{figure}

Using the asymptotic formula \eqref{Pnt-1} and replacing summation
with integration, we find
\begin{eqnarray}
\label{Q:asymp}
\mathbb{P}_n &\simeq& \frac{8}{\sqrt{2\pi t}}\sum_{j\geq 1}(-1)^{j-1} j^2\int_n^\infty dk\,\exp\!\left[-\tfrac{j^2 k^2}{2t}\right]  \nonumber \\
&=& 4 \sum_{j\geq 1}(-1)^{j-1} j\,\text{Erfc}\left(\frac{j\,n}{\sqrt{2t}}\right)
\end{eqnarray}
where $\text{Erfc}(y)=\frac{2}{\sqrt{\pi}}\int_y^\infty dz\,e^{-z^2}$
is the error function.  Differentiating \eqref{Q:asymp} yields
\begin{equation}
\label{sum-tie:1}
\frac{d\mathbb{P}_n}{dt}  \simeq t^{-1}\,\frac{4}{\sqrt{\pi}}\sum_{j\geq 1}(-1)^{j-1} j^2 \nu\, e^{-j^2 \nu^2}
\end{equation}
Substituting \eqref{Pnt-1} and \eqref{sum-tie:1} into \eqref{nu-PQ}
and replacing summation over $n$ with integration over
$\nu=n/\sqrt{2t}$ we arrive at $\frac{dA}{dt} \simeq\frac{a_1}{t}$ [and hence $A(t)\simeq a_1 \ln t$] with
\begin{equation}
\label{a1-int-sum}
a_1= \frac{64}{\pi} \int_0^\infty d\nu\,\sum_{i, j\geq 1}(-1)^{i+j} i^2  j^2 \nu\, e^{-(i^2+j^2) \nu^2}
\end{equation}
It is possible to simplify the integral over the double sum into a
compact sum (Appendix~\ref{app:A1})
\begin{equation}
\label{a1-simple}
a_1=  16\sum_{ j\geq 1}\frac{(-1)^{j-1}\,  j^3}{\sinh(\pi j)}=0.970\,508\,\ldots
\end{equation}
The simulation results are in excellent agreement with this
theoretical prediction: The numerically measured amplitude,
$a_1=0.970\pm 0.001$, is within $0.05\%$ of the theoretical value (see
also Fig.~\ref{fig-1d-nu}).

When $d\geq 2$, the range is a self-averaging quantity with the
asymptotic distribution
\begin{equation}
\label{P-sigma}
P_n(t)\simeq \frac{1}{\sqrt{V}}\,\mathcal{P}_d(\sigma), \qquad \sigma=\frac{n-N}{\sqrt{V}}
\end{equation}
By inserting \eqref{P-sigma} into \eqref{nu-PQ} we obtain
\begin{eqnarray}
\label{nu-2+}
\frac{dA}{dt}  \simeq  \frac{2}{\sqrt{V}}\,\frac{d N}{dt} \int_{-\infty}^\infty d\sigma\,[\mathcal{P}_d(\sigma)]^2
\end{eqnarray}
to leading order. 

\begin{figure}[t]
\begin{center}
\includegraphics[width=0.425\textwidth]{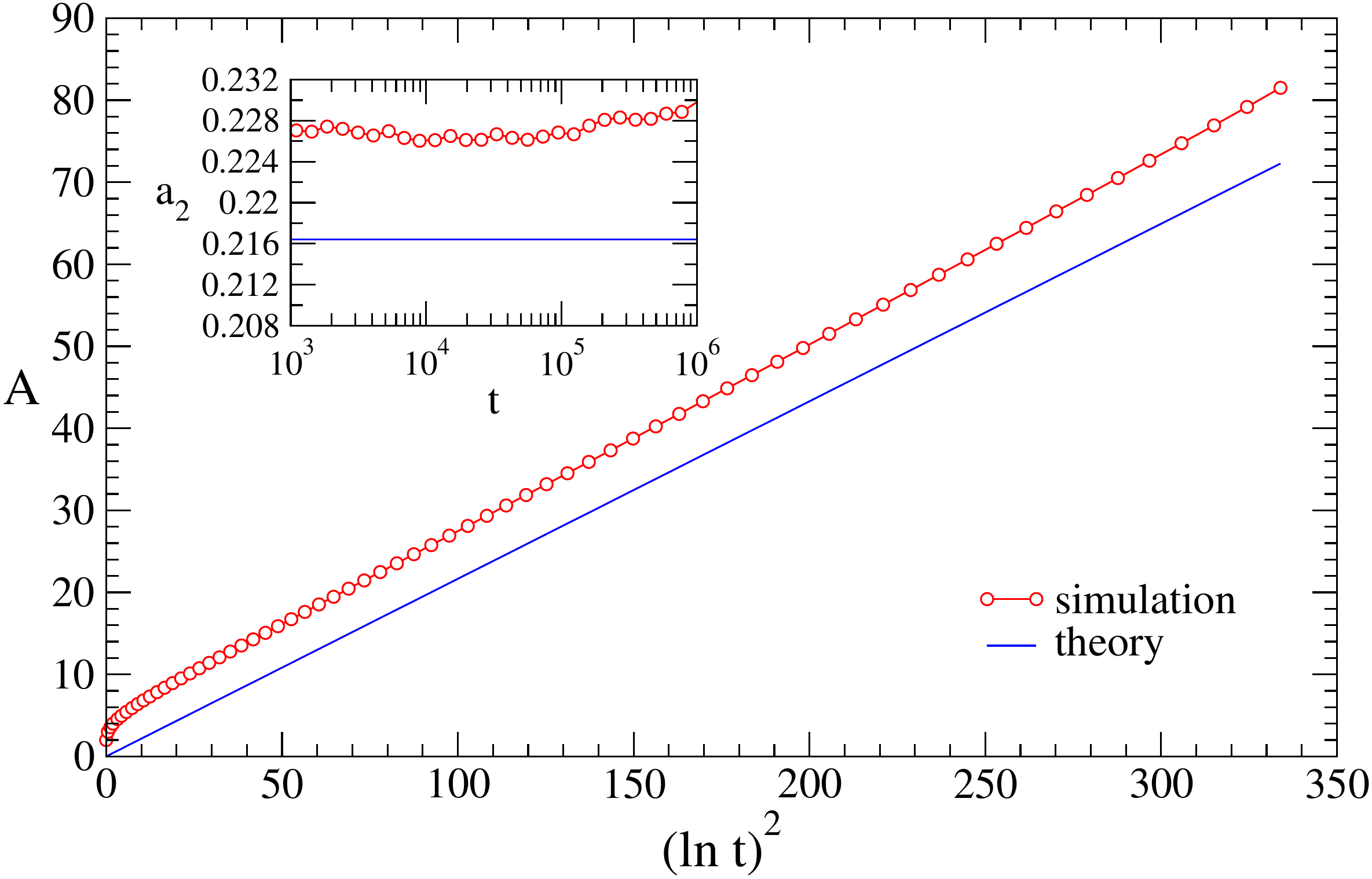}
\caption{The average number of ties $A$ versus $(\ln t)^2$ in two
  dimensions. Simulation results are compared with the theoretical
  prediction. The inset compares simulation results for the pre-factor
  $a_2(t) \equiv dA/d(\ln t)^2$ with the approximate value
  \eqref{A-2-Gauss}. An average over $2^{16}$ independent runs has
  been performed.}
\label{fig-2d-nu}
  \end{center}
\end{figure}

In two dimensions, $\frac{d N}{dt} \simeq \frac{\pi}{\ln t}$ and
\hbox{$V\simeq V_2 t^2/(\ln t)^4$}, and therefore, Eq.~\eqref{nu-2+}
leads to the asymptotic behavior \hbox{$A(t)\simeq a_2\,(\ln t)^2$}
in \eqref{ties-d} with
\begin{equation}
\label{A-2}
a_2 = \frac{\pi}{\sqrt{V_2}}\int_{-\infty}^\infty d\sigma\,[\mathcal{P}_2(\sigma)]^2
\end{equation}
The range distribution $\mathcal{P}_2$ is not Gaussian in two
dimensions, but it has been probed numerically in Ref.~\cite{Wiener00}
and was found to be close to Gaussian~\footnote{Rather than the random
  walk on the square lattice, a Wiener sausage in the plane was
  studied in \cite{Wiener00}. A Wiener sausage is a fattened Brownian
  trajectory; the standard Wiener sausage is a domain covered by a
  spherical Brownian particle. The volume of the Wiener sausage is
  analogous of the range of the random walk.}.  Substituting
$\mathcal{P}_2^\text{Gauss}=(2\pi)^{-1/2}e^{-\sigma^2/2}$ into
\eqref{A-2} yields the uncontrolled approximation (see
Fig.~\ref{fig-2d-nu})
\begin{equation}
\label{A-2-Gauss}
a_2^\text{Gauss} = \sqrt{\frac{\pi}{4V_2}}=0.216\ldots
\end{equation}
Numerically we measured \hbox{$a_2 = 0.227 \pm
  0.001$}, a value within 5\% of \eqref{A-2-Gauss}. Thus, the
uncontrolled Gaussian approximation yields a close estimate for the
amplitude $a_2$. 

\begin{figure}[t]
\begin{center}
\includegraphics[width=0.425\textwidth]{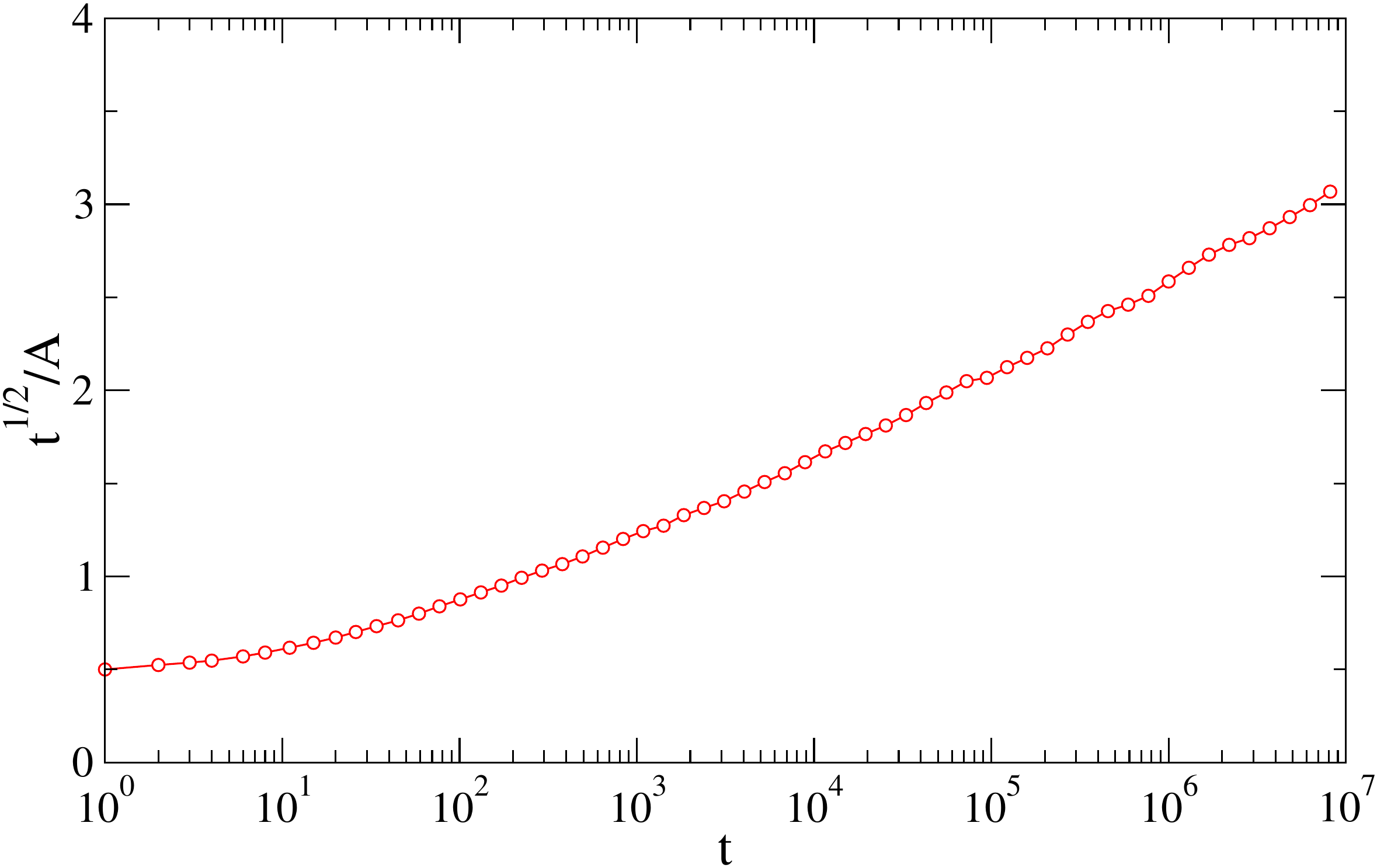}
\caption{The quantity $\sqrt{t}/A$ versus time in three dimensions.}
\label{fig-At-3d}
  \end{center}
\end{figure}

When $d=3$, the range distribution is Gaussian. Using
$\mathcal{P}_3=(2\pi)^{-1/2}e^{-\sigma^2/2}$, $\frac{d N}{dt} \simeq
\frac{1}{W_3}$ and $V=V_3\, t\ln t$ we recast Eq.~\eqref{nu-2+} into
\begin{eqnarray}
\label{nu-t-3}
\frac{dA}{dt} \simeq   \frac{1}{W_3\sqrt{\pi V_3\, t\ln t}}
\end{eqnarray}
Performing the integration yields $A(t)\simeq a_3\sqrt{t/\ln
  t}$ with
\begin{equation}
\label{A-3}
a_3 = \frac{2}{W_3 \sqrt{\pi V_3}}= \frac{\sqrt{\pi}\,W_3}{(W_3-1)^2}= 10.079\,423\ldots
\end{equation}
However, numerically we find that $A \sim \sqrt{t}/\ln t$ provides a
significantly better fit to simulation results than the theoretical
prediction $A \sim \sqrt{t/\ln t}$, see Fig.~\ref{fig-At-3d}.

When $d\geq 4$, Eq.~\eqref{nu-2+} becomes 
\begin{eqnarray}
\label{nu-t-d}
\frac{dA}{dt} \simeq   \frac{1}{W_d\sqrt{\pi V_d t}}
\end{eqnarray}
leading to $A(t)\simeq a_d\sqrt{t}$ as in \eqref{ties-d} with the
amplitude 
\begin{equation}
\label{A-345}
a_d = \frac{2}{W_d \sqrt{\pi V_d}}
\end{equation}
This completes the derivation of the asymptotic behaviors
\eqref{ties-d} with the amplitudes \eqref{a1-simple}, \eqref{A-3}, and
\eqref{A-345}. However, the numerical simulation results are at odds
with the theoretical prediction for $a_4$. Numerically, we measured
$W_4=1.24\pm 0.01$ in agreement with $W_4=1.239467$ \cite{Griffin}
that follows from \eqref{Watson}.  Numerical simulations yield the
amplitude $V_4=0.26\pm 0.01$ for the variance. Accordingly,
Eq.~\eqref{A-345} gives $a_4=1.78$ but the numerical simulations yield
$a_4=1.04\pm 0.01$.

The diffusive growth $A(t) \sim \sqrt{t}$ for $d\geq 4$, which we 
verified numerically for $d= 4$, can be deduced using heuristic
arguments.  In the limit $d\to\infty$, the quantities $\mathcal{N}_1$
and $\mathcal{N}_2$ become Markovian.  These two quantities reduce to
directed random walks: each directed walk undergoes $+1$ hops with
unit rate. Hence, the difference $\mathcal{N}_1-\mathcal{N}_2$
performs a one-dimensional symmetric random walk as it undergoes $\pm
1$ jumps, both with unit rate. Consequently, the number of ties is
equivalent to the number of times a symmetric random walk returns to
the origin.  As a result, the average number of ties grows
diffusively, $A(t) \sim \sqrt{t}$, in the limit $d\to\infty$. While
this diffusive growth is formally justified only in infinite
dimension, we expect this behavior to hold for all $d\geq 4$. 

The logarithmic growth of the number of ties in one dimension
resembles the growth law \eqref{ties-M} corresponding to ties between
maxima.  The probability to observe $n$ ties between maxima of two
random walks during the time interval $(0,t)$ was found to be
Poissonian $\sim t^{-1/4}(\ln t)^n$ \cite{BK-maxima-leads}.  We
anticipate a similar functional form holds for ties between the ranges
of two random walks,
\begin{equation}
\label{Phi-n}
\Phi_n(t) \sim t^{-2/3}(\ln t)^n
\end{equation}
Numerically, we confirmed \eqref{Phi-n} for $n=0,1,2,3$.

\section{Multiple random walks}
\label{sec:many}

The probability $P_m(t)$ that the ranges of $m$ random walks remain
perfectly ordered till time $t$, defined by
\begin{equation}
\label{Pt-m:def}
P_m(t)  = \text{Prob}[\mathcal{N}_1(\tau)\leq \cdots\leq  \mathcal{N}_m(\tau)\,|\,0\leq \tau\leq t] 
\end{equation}
is a straightforward generalization of \eqref{Pt:def}. We compare this
quantity with the probability 
\begin{equation}
\label{ord-m:def}
\Pi_m(t) = \text{Prob}[x_1(\tau)\leq \cdots\leq   x_m(\tau)\,|\,0\leq \tau\leq t]
\end{equation}
that the {\em positions} of $m$ one-dimensional random walks remain ordered till time $t$,
When $t\to\infty$, these ordering probabilities decay algebraically with time
\cite{Karlin59,Fisher84,Grabiner99}, 
\begin{equation}
\label{ord-m:decay}
\Pi_m(t) \sim t^{-\overline{\beta}_m}\,, \qquad     \overline{\beta}_m = \tfrac{1}{4}m(m-1)
\end{equation}

In high dimensions, the range of a random walk undergoes a
one-dimensional directed random walk.  Hence, the asymptotic behavior
of the ordering probability $P_m(t)$ when $d\geq 4$ is specified in
Eq.~\eqref{ord-m:decay}. Based on the asymptotic behaviors of $P_2(t)$
given in Eqs.~\eqref{Pt-d-precise}, we conjecture 
\begin{equation}
\label{Pt-d-m}
P_m \sim
\begin{cases}
t^{-\beta_m}                              & d=1\\
t^{-\beta_m}  (\ln t)^{h_m}         & d=2\\
t^{-m(m-1)/4}(\ln t)^{-g_m}        & d= 3\\
t^{-m(m-1)/4}                             & d\geq 4
\end{cases}
\end{equation}
The set of algebraic exponents $\beta_m$ characterizes $P_m(t)$ in one
dimension (see Table I), and additionally, the logarithmic exponents
$h_m$ and $g_m$ characterize this ordering probability when $d=2$ and $d=3$.

We also studied the probability $L_m(t)$ that the range of one walk
(the leader) exceeds that of every other walk during the time interval
$(0,t)$, that is, 
\begin{equation*}
\label{Lt-m:def}
L_m(t)  = \text{Prob}[\mathcal{N}_1(\tau)\geq \mathcal{N}_j(\tau)\,|\,j=2,\ldots, m; 0\leq \tau\leq t] 
\end{equation*}
Numerically, we find that the ordering probability 
exhibits an  algebraic decay in one dimension (see Table I)
\begin{equation}
\label{alpha-beta:m}
L_m \sim t^{-\alpha_m}
\end{equation}
This algebraic decay is similar to that of the ordering probability
$P_m(t)$, see Eq.~\eqref{Pt-d-m}. In Table~\ref{Table:alpha-beta}, we
also list the set of exponents $\overline{\alpha}_m$ characterizing
the decay of the probability that a random walker remains in the
lead position \cite{Bramson,KR96a,KR96b,KR99,Avraham03,BK-cone,BK-multiple},
that is
\begin{equation*}
\label{xt-m:def}
\text{Prob}[x_1(\tau)\geq x_j(\tau)\,|\,j=2,\ldots, m, ~0\leq \tau\leq t] \sim t^{-\overline{\alpha}_m}
\end{equation*}

The exponents presented in Table~\ref{Table:alpha-beta} indicate
\hbox{$\alpha_m>\overline{\alpha}_m$} and
\hbox{$\beta_m>\overline{\beta}_m$} for all $m\geq 2$. The growth of
the exponents $\alpha_m$ and $\beta_m$ with $m$ resembles that 
of $\overline{\alpha}_m$ and $\overline{\beta}_m$. The asymptotic
growth is $\overline{\alpha}_m\simeq \frac{1}{4}\ln m$ for $m\gg 1$,
see \cite{KR96a,KR96b,KR99}.

Using heuristic arguments, it is possible to show that the leading
large-$m$ behaviors of $\alpha_m$ and $\overline{\alpha}_m$ are the
same.  First, we recall the known derivation for the quantity
$\overline{\alpha}_m$. The boundaries of the region visited by random
walks other than the leader become more and more deterministic as
$m\to\infty$. The region is asymptotically symmetric with respect to
the origin, $(-x_*, x_*)$, with $x_*$ estimated from the criterion
\begin{equation}
\label{crit-x}
\int_{x_*}^\infty dx\,\frac{m-1}{\sqrt{2\pi t}}\,e^{-x^2/2t} \sim 1
\end{equation}
An elementary asymptotic analysis yields
\begin{equation}
\label{x-star}
x_*\simeq \sqrt{2Ct}, \qquad C = \ln m
\end{equation}
The leader must stay in the region $x>x_*=\sqrt{C\tau}$
during the time interval $0<\tau<t$. This problem admits an exact
solution \cite{KR96a,KR96b,KR99} for arbitrary $C>0$, namely, the
survival probability decays as $t^{-\alpha}$ with
$\alpha=\alpha(C)$. We are interested in the $m\gg 1$ behavior, and
generally, the deterministic description of the boundaries is
asymptotically exact only when $m\gg 1$. Thus $C\gg 1$, and in this
situation $\alpha\simeq (C-1)/4$, see \cite{KR96a,KR96b,KR99}, with
our choice of the diffusion coefficient $D=\frac{1}{2}$. From
\eqref{crit-x} we obtain $C\simeq \ln m$. Thus, we recover
$\overline{\alpha}_m\simeq \frac{1}{4}\ln m$ for $m\gg 1$.

\begin{table}[t]
\begin{tabular}{|c|l|l|l|l|l|l|}
\hline
$m$                              &  ~~2  &  ~~3     &   ~~4    &~~ 5     & ~~6     & ~~~$m \gg 1$  \\
\hline
$\alpha_m$                  &0.667 & 0.947   &\,1.103   &1.233   &1.315   &  ~~$(\ln m)/4$\  \\
\hline
$\overline{\alpha}_m$  &~1/2   & ~3/4    &0.9134   &\,1.03    &\,1.11   &  ~~$(\ln m)/4$\\
\hline
$\beta_m$                    &0.667 &  ~1.91  & ~3.65   & ~~6.0   & ~~8.3 &  ~~~~$Bm^2$ \\
\hline
$\overline{\beta}_m$    &~1/2   & ~3/2      &~~~3    &~~~5    &\,15/2   &  $m(m-1)/4$
\\
\hline
\end{tabular}
\caption{The exponents $\alpha_m$ and $\beta_m$ obtained from
  numerical simulations of $m$ one-dimensional random walks for
  $m=2,\ldots,6$. The exponents $\overline{\alpha}_m$ and
  $\overline{\beta}_m$ characterizing similar ordering of the
  positions of one-dimensional random walks are listed as a reference.
  The exponents $\overline{\alpha}_2=\frac{1}{2}$,
  $\overline{\alpha}_3=\frac{3}{4}$ are known analytically.  The
  leading large $m$ behaviors is shown in the last column. The
  asymptotic behavior $\beta_m\simeq Bm^2$ is conjectural, the
  amplitude $B$ is unknown.}
\label{Table:alpha-beta}
\end{table}

The range distribution \eqref {Pnt-1} simplifies to 
\begin{equation}
\label{Pnt-1-asym}
P_n(t)\simeq \frac{8}{\sqrt{2\pi t}}\,\exp\!\left[-\frac{n^2}{2t}\right]
\end{equation}
in the limit $n\gg \sqrt{t}$. The probability of finding a random walk
of range $n$ and other $m-2$ random walks of smaller range is
\begin{equation}
(m-1)P_n (1-\mathbb{P}_n)^{m-2}\simeq (m-1)P_n e^{-(m-2)P_n}
\end{equation}
with $P_n$ given by  \eqref {Pnt-1-asym}. The criterion
\begin{equation}
\label{crit-n}
\sum_{n\geq  n_*} (m-1)P_n e^{-(m-2)P_n}\sim 1
\end{equation}
gives $n_*\simeq \sqrt{2Ct}$ and the same $C\simeq \ln m$ in the
leading order. The range is anomalously large, so the leader must stay
in the region $x>x_*=n_*=\sqrt{C\tau}$. Thus, we arrive at the same
leading behavior $\alpha_m\simeq \frac{1}{4}\ln m$.

We have also investigated the average number of distinct complete ties for three random walks, i.e., instances when $\mathcal{N}_1=\mathcal{N}_2=\mathcal{N}_3$. Using the rate equation approach, we have found that the number of ties saturates at a finite value when $d\leq 2$, while for $d\geq 3$, it grows indefinitely with time. Numerical simulations confirm these qualitative behaviors. For $m\geq 4$ random walks, the number of complete ties remains finite in any dimension. 

\section{Discussion}
\label{sec:disc}

We investigated the competition between sets visited by two identical
random walks on hyper-cubic lattices. We also studied the race between
the ranges of the walks.  Using analytic methods, we studied the
asymptotic behaviors \eqref{ties-d} for the average number of ties
between the ranges of the two walks.  We found that the average number
of ties grows as $\ln t$ in one dimension and as $(\ln t)^2$ in two
dimensions.

We also studied ordering probabilities associated with the number of
sites and the set of visited sites. In general, the ordering
probabilities decay algebraically in one dimension, and a challenge
for future work is an analytic determination of the decay exponents
$\beta$ and $\gamma$. The behavior of the ordering probabilities in
higher dimensions is much richer. Of special interest is the ordering
probability $Q(t)$ associated with the sets of visited sites, viz.,
the probability that the set of sites visited by one random walker
remains a subset of the sites visited by another. Numerically, we find
that the ordering probability $Q(t)$ decays as a stretched exponential
in $d\geq 2$.  Determining the quantity $Q(t)$ analytically is a
formidable challenge

The probabilities $P(t)$ and $Q(t)$ may be also studied for Brownian
motions \cite{IM:book,BM:book}.  To have
well-defined ordering probabilities, we postulate that $S_1(0)\subset
S_2(0)$ , e.g., $S_2(0)=[-\epsilon,\epsilon]$ and $S_1(0)$ is the
origin. The ranges $|S_j(t)|$ are now positive real numbers, and the
probabilities $P(t)$ and $Q(t)$ are well defined by \eqref{Pt:def} and
\eqref{Ot:def}.  The decay laws \eqref{alpha} and \eqref{gamma}
acquire dimensionally consistent form
\begin{equation}
\label{PP:BM}
P(t) \sim \left(\frac{\epsilon^2}{Dt}\right)^{\beta}\,, \qquad 
Q(t)  \sim  \left(\frac{\epsilon^2}{Dt}\right)^\gamma
\end{equation}
where $D$ is the diffusion coefficient. For Brownian motion in $d\geq 
2$, one can consider a Wiener sausage containing all points within a
fixed distance from the Brownian trajectory, i.e., a domain visited by
a spherical Brownian particle. For Wiener sausages, the average
visited volume and its variance are thoroughly understood \cite{LeGall88,Wiener89,Wiener01,Wiener04,Wiener21,Hamana10,Hamana16,Asselah19a}, and
qualitatively similar to \eqref{Nt-d} and \eqref{Vt-d}. 

In this study, we addressed the probability that one random walk never
visits a site that was not previously visited by another walk. A
complementary and related question involves non-intersection probabilities \cite{Lawler82,Duplantier88a,Chen10} describing
realizations when trajectories do not intersect, i.e., any two walks never visit the same site.
Conformal field theory, two-dimensional quantum gravity, and
Schramm-L\"{o}wner evolution have been applied
\cite{Duplantier88b,Duplantier98,LSW1,LSW2,Duplantier03,Lawler05,Duplantier06}
to study non-intersection probabilities in two dimensions. Methods
used for the analysis of non-intersection probabilities in higher dimensions \cite{Chen10,Lawler82,Duplantier88a} could perhaps 
be adapted to the analysis of the ordering probabilities.

In addition to the average number of ties between the ranges of two random walks, one can study further statistical properties of the number of ties. Another natural direction for future work is to investigate ties between the sets of visited sites.

\appendix
\vspace{-0.3in}
\section{Derivation of Eq.~\eqref{a1-simple}}
\label{app:A1}

To simplify the right-hand side of \eqref{a1-int-sum} let us reverse
the order of summation and integration, i.e., first integrate term by
term. The sum in
\begin{equation}
\label{a1-double-sum}
a_1= \frac{32}{\pi} \sum_{i, j\geq 1}(-1)^{i+j}\, \frac{ i^2  j^2 }{i^2+j^2}
\end{equation}
is formally divergent. Regularization allows us to deduce a
finite answer. Rearranging the terms in the sum yields
\begin{eqnarray}
\label{massage}
\sum_{i, j\geq 1}(-1)^{i+j}\, \frac{ i^2  j^2 }{i^2+j^2} &=& \sum_{i, j\geq 1}(-1)^{i+j} j^2\,\frac{ i^2 + j^2 - j^2}{i^2+j^2} \nonumber\\
&=& \sum_{i, j\geq 1}(-1)^{i+j} \left[j^2-\frac{ j^4}{i^2+j^2}\right]  \nonumber \\
&=& -\sum_{i, j\geq 1}(-1)^{i+j} \frac{ j^4}{i^2+j^2}
\end{eqnarray}

In the last step, we used \hbox{$\sum_{j\geq 1}(-1)^{j} j^2 =
  0$}. Indeed, the sum is equal to $7\zeta(-2)=0$, with zeta function
$\zeta(s)=\sum_{j\geq 1}j^{-s}$ at $s=-2$ viewed as an analytic
continuation of $\zeta(s)$ defined when $\text{Re}(s)>1$.  The zeta
function vanishes at all even negative integers,
$\zeta(-2p)=\sum_{j\geq 1} j^{2p}=0$, as discovered by Euler, see
\cite{Kato,Varadarajan07}.

We now perform the summation in \eqref{massage} over $i\geq 1$ using the identity
\begin{equation}
\label{identity}
\sum_{i\geq 1}\frac{(-1)^i}{i^2+j^2}=\frac{\frac{\pi j}{\sinh(\pi j)}-1}{2j^2}
\end{equation}
Inserting \eqref{massage}  and \eqref{identity} into \eqref{a1-double-sum} 
we obtain 
\begin{equation}
\label{a1-sum}
a_1= \frac{16}{\pi} \sum_{ j\geq 1}(-1)^{j}\left[j^2-\frac{\pi j^3}{\sinh(\pi j)}\right]
\end{equation}
Using the identity \hbox{$\sum_{j\geq 1}(-1)^{j} j^2 = 0$} again we
simplify \eqref{a1-sum} to Eq.~\eqref{a1-simple}.

\bibliography{references-RW}

\end{document}